\documentclass[twocolumn, twocolappendix]{aastex631}

\usepackage{amsmath}
\usepackage{amssymb}
\usepackage{graphicx}
\usepackage{bm}
\usepackage{soul}

\newcommand{\fNLloc}{f_\mathrm{NL}^\mathrm{local}}
\newcommand{\fNLeq}{f_\mathrm{NL}^\mathrm{equil}}
\newcommand{\fNLort}{f_\mathrm{NL}^\mathrm{ortho}}

\newcommand{\Om}{\Omega_\mathrm{m}}
\newcommand{\Mmin}{M_\mathrm{min}}

\newcommand{\hMpc}{h\,\mathrm{Mpc}^{-1}}

\newcommand{\kmax}{k_\mathrm{max}}

\def\btheta{\boldsymbol{\theta}}
\def\bs{\mathbf{s}}
\def\cov{\mathbf{C}}

\newcommand{\Planck}{\textit{Planck}}
\newcommand{\Quijote}{\textsc{Quijote}}
\newcommand{\QuijotePNG}{\textsc{Quijote-png}}

\newcommand{\ias}{Universit\'{e} Paris-Saclay, CNRS, Institut d’Astrophysique Spatiale, 91405, Orsay, France}
\newcommand{\iap}{Sorbonne Universit\'{e}, CNRS, UMR 7095, Institut d'Astrophysique de Paris, 98 bis bd Arago, 75014 Paris, France}
\newcommand{\cca}{Center for Computational Astrophysics, Flatiron Institute, 162 5th Avenue, New York, NY 10010, USA}
\newcommand{\bologna}{Dipartimento di Fisica e Astronomia, Alma Mater Studiorum - University of Bologna, Via Piero Gobetti 93/2, 40129 Bologna BO, Italy}
\newcommand{\inaf}{INAF - Osservatorio Astronomico di Bologna, Via Piero Gobetti 93/3, 40129 Bologna BO, Italy}
\newcommand{\infn}{INFN - Istituto Nazionale di Fisica Nucleare, Sezione di Bologna, Viale Berti Pichat 6/2, 40127 Bologna BO, Italy
}
\newcommand{\infnPad}{INFN, Sezione di Padova, via Marzolo 8, I-35131, Padova, Italy}
\newcommand{\Trento}{Dipartimento di Fisica, Universit\`a degli Studi di Trento, Via Sommarive 14, I-38123 Povo (TN), Italy}
\newcommand{\ICC}{ICC, University of Barcelona, IEEC-UB, Martí i Franquès, 1, E-08028 Barcelona, Spain}
\newcommand{\Galilei}{Dipartimento di Fisica e Astronomia “G. Galilei”,Università degli Studi di Padova, via Marzolo 8, I-35131, Padova, Italy}
\newcommand{\princeton}{Department of Astrophysical Sciences, Princeton University, 4 Ivy Lane, Princeton, NJ 08544 USA}
\newcommand{\Kavli}{Kavli Institute for Cosmology Cambridge, Madingley Road, Cambridge CB3 0HA, UK}
\newcommand{\DAMTP}{DAMTP, Centre for Mathematical Sciences, University of Cambridge, Wilberforce Road, Cambridge CB3 OWA, UK}

\newcommand{\equalContributions}{Equal contributions.}

\begin{document}
\title{Quijote-PNG: Optimizing the summary statistics to measure Primordial non-Gaussianity}

\author{Gabriel Jung} \altaffiliation{\equalContributions} \affiliation{\ias}
\author{Andrea Ravenni}\altaffiliation{\equalContributions} \affiliation{\Galilei} \affiliation{\infnPad} \affiliation{\ICC}
\author{Michele Liguori} \affiliation{\Galilei} \affiliation{\infnPad} \affiliation{\Trento}
\author{Marco Baldi}\affiliation{\bologna}\affiliation{\inaf}\affiliation{\infn}
\author{William R. Coulton}\affiliation{\Kavli}\affiliation{\DAMTP}
\author{Francisco Villaescusa-Navarro} \affiliation{\cca} \affiliation{\princeton}
\author{Benjamin D. Wandelt}\affiliation{\iap}\affiliation{\cca}

\begin{abstract}
We apply a suite of different estimators to the \QuijotePNG\ halo catalogues to find the best approach to constrain Primordial non-Gaussianity (PNG) at non-linear cosmological scales, up to $k_{\rm max} = 0.5 \, h\,{\rm Mpc}^{-1}$. 
The set of summary statistics considered in our analysis includes the power spectrum, bispectrum, halo mass function, marked power spectrum, and marked modal bispectrum. Marked statistics are used here for the first time in the context of PNG study. We perform a Fisher analysis to estimate their cosmological information content, showing substantial improvements when marked observables are added to the analysis. Starting from these summaries, we train deep neural networks (NN) to perform likelihood-free inference of cosmological and PNG parameters.
We assess the performance of different subsets of summary statistics;
in the case of $f_\mathrm{NL}^\mathrm{equil}$, we find that a combination of the power spectrum and a suitable marked power spectrum outperforms the combination of power spectrum and bispectrum, the baseline statistics usually employed in PNG analysis.
A minimal pipeline to analyse the statistics we identified can be implemented either with our ML algorithm or via more traditional estimators, if these are deemed more reliable.
\end{abstract}

\section{Introduction}
\label{sec:introduction}

Primordial non-Gaussianity (PNG) provides a potentially powerful tool to discriminate between different Early Universe scenarios and its investigation plays therefore an important role in observational cosmology. 

However, extracting PNG information is a task made significantly difficult by the smallness of the expected signal, which at low redshifts is several orders of magnitude below that generated by non-linear gravitational evolution of cosmological perturbations 
For this reason, most observational studies of primordial non-Gaussian parameters have so far focused on linear cosmological probes, such as the Cosmic Microwave Background (CMB) \citep{Planck:2019kim}, or the galaxy power spectrum and bispectrum on large scales \citep{Cabass:2022wjy, Cabass:2022ymb, DAmico:2022gki, Cagliari:2023mkq, Ivanov:2024hgq}.

While challenging, as we have just stressed, a Large Scale Structure (LSS) analysis at non-linear scales could potentially be very rewarding: most of the PNG constraining power comes in fact from the cosmological perturbation bispectrum and a simple mode counting argument suggests that large improvements could be achieved in this regime, provided we are able to at least partially clean the total non-Gaussianity
signal from late-time non-linear contributions.
In this work---which is connected to a series of previous studies in the \QuijotePNG\ series \citep{Coulton:2022rir, Jung:2022gfa, Jung:2023kjh}---we investigate this possibility by performing a thorough analysis of the dark matter halo field in N-body simulations with PNG initial conditions, testing scales up to $k_{\rm max} = 0.5 \, {h\, {\rm Mpc}^{-1}}$. 

The study of cosmological information at non-linear scales, in a more general context than just PNG analysis, is actually a research line that is recently receiving considerable attention. The reason for this growing interest is likely twofold. On one side, a large amount of data from coming galaxy surveys is going to make small scales observably accessible with high precision. 
On the other side, the past few years have seen significant methodological developments in cosmological data analysis, like field-level inference, whether it is performed using perturbative \citep{Schmidt:2020ovm, Baumann:2021ykm, Cabass:2023nyo} or Bayesian \citep{Jasche:2012kq, Andrews:2022nvv} forward models, or via machine learning using simulations \citep{Ribli:2019wtw, Ntampaka:2019ole, Villanueva-Domingo:2022rvn, Makinen:2022jsc, Shao:2022mzk, deSanti:2023zzn, Lemos:2023myd, Roncoli:2023ysi}. This kind of simulation-based approach is also explored with data compressed into suitable sets of summary statistics \citep{Alsing:2018eau, Alsing:2019xrx, Jeffrey:2020xve, Hahn:2022wgo, Hahn:2022zxa, Blancard:2023iab, Hahn:2023kky, Hahn:2023udg, Hou:2024blc, DES:2023qwe, Tucci:2023bag, Ivanov:2024hgq}.

Here, we take the latter approach to the problem, first compressing the data into a set of pre-determined summary statistics. While potentially leading to some loss of information with respect to a full field level analysis, this procedure presents some advantages, as it is less computationally demanding, potentially easier to implement when analyzing actual data and, above all, it generally leads to results that are more amenable to a clear physical interpretation. The feature of interpretability may also aid in separating effects due to systematics from those due to the signal under study.
Beyond quantifying the amount of information on PNG contained in different summary statistics computed from the halo density field at late times, the main goal of our work is to determine a suitable subset, which reaches the optimal compromise between being informative and relatively easy to analyse.

In previous works, we started pursuing this program by considering different combinations of the halo power spectrum, bispectrum and halo mass function. Here, we start by extending this analysis with the inclusion of additional summaries, the marked power spectrum and marked bispectrum. These marked statistics provide a flexible and easy way to measure weighted combinations of n-point correlation functions and previous studies showed that they are able to tightly constrain neutrino masses at non-linear scales \citep{Massara:2020pli, Philcox:2020fqx, Massara:2022zrf}. They are therefore a natural option to consider also in a PNG analysis and we will indeed show in this paper that they can provide a powerful tool to constrain PNG parameters.

Besides adding new summaries, the present work also contains some significant methodological extensions of our previous studies. In previous works \citep{Coulton:2022rir, Jung:2022gfa, Jung:2023kjh}, we used the Fisher matrix formalism to provide figures of merit for the various statistics. These were calculated, at fully non-linear scales for a fiducial cosmology, by numerically evaluating derivatives and covariance matrices through Monte Carlo averaging of tens of thousands of N-body realizations of the halo field.
The Fisher matrices---obtained following the algorithm by 
\citet{Coulton:2023sfu}
based on score compression into a minimal set of summaries---were also used to build and test quadratic estimators, which are nearly optimal for parameter values near the fiducial cosmology.
One drawback of this approach is that it makes the exploration of a wide range of parameter values very computationally demanding, as it requires to produce new sets of tens of thousand simulations, for many different choices of fiducial parameters.
We try in this work to overcome this limitation by using deep neural networks (NN), trained on a suite of simulations whose parameters are arranged in a Latin-hypercube, to map our summaries directly into the final parameters, without having to explicitly evaluate any covariance or make any assumption about the likelihood function.
We discuss a first application of our pipeline, mainly aimed at a detailed comparison of many NG statistics, in order to find their optimal combination in terms of PNG sensitivity and simplicity of implementation.
This is a first step in the direction of future applications to PNG parameter inference on real data from galaxy surveys, which will require the use of realistic galaxy mocks.

The plan of the paper is as follows. In Section~\ref{sec:simulations} we briefly describe the \QuijotePNG\ simulation suite that was used to calculate the Fisher matrices, train the networks and produce the final forecasts; in Section~\ref{sec:statistics} we introduce the summary statistics considered in our analysis and discuss the methods used to extract them from the data; in Section~\ref{sec:method} we discuss the implementation of the NNs and the metrics that we use to assess their performance; in Section~\ref{sec:analysis} we show our results, which include Fisher forecasts and a comparison of different data pre-processing methodologies, followed by NN results for many different combinations of summaries; in Section~\ref{sec:conclusion} we draw our final conclusions.

\newpage
\section{Simulations}
\label{sec:simulations}

The analyses presented in this paper are based on the \Quijote\ \citep{Villaescusa-Navarro:2019bje} and \QuijotePNG\ \citep{Coulton:2022qbc} sets of simulations.

These are dark matter only N-body simulations of volume $1\left( h^{-1}{\rm Gpc} \right)^3$, containing $512^3$ particles each, and run using the TreePM code \textsc{Gadget-III} from initial conditions generated at $z=127$ by the codes \textsc{2LPTIC} \citep{Crocce:2006ve} and \textsc{2LPTPNG} \citep{Scoccimarro:2011pz, Coulton:2022qbc}\footnote{\url{https://github.com/dsjamieson/2LPTPNG}}, for the simulations without and with PNG, respectively. We focus on dark matter halos, which are identified in each simulation by the standard Friends-of-friends algorithm \citep{1985ApJ...292..371D} by setting the linking length parameter to $b=0.2$ and considering halos with more than 20 dark matter particles. 

We mainly use a set of $1000$ simulations with varying amount of equilateral PNG, with $\fNLeq \in [-600,600]$, and varying cosmological parameters, where the parameters are distributed in a Latin-hypercube (LH). We also work with the $15000$ \Quijote\ simulations at a fiducial cosmology compatible with \Planck\ CMB observations and without PNG and, to compute Fisher forecasts and compressed statistics, we use additional sets of $500$ simulations, in which one parameter has been slightly displaced with respect to its fiducial value. We perform extra tests on the original \Quijote\ LH ($2000$ simulations with varying cosmological parameters and no PNG) and a LH with fixed cosmological parameters and varying local PNG, with $\fNLloc \in [-300,300]$. The main characteristics of all these simulations are given in Table~\ref{tab:quijote}. We release the local and equilateral LH in complement to this work, which makes all simulations used here publicly available.\footnote{\url{https://quijote-simulations.readthedocs.io/en/latest/png.html}}

\begin{deluxetable*}{c|c|ccccc|ccc}
\tablecaption{Parameters of the \Quijote~and \QuijotePNG~halo catalogues used in this work.  \label{tab:quijote}}
\tablehead{& $N_\mathrm{sims}$ & $\sigma_8$ & $\Om$ & $\Omega_{\rm b}$ & $n_s$ & $h$ & $\fNLloc$ & $\fNLeq$ & $\fNLort$}
\startdata
\textbf{Fiducial} & 15000 & 0.834 & 0.3175 & 0.049 & 0.9624 & 0.6711 & 0 & 0 & 0 \\
\textbf{Displaced} & 500 & $\pm$ 0.015 & $\pm$ 0.01 & $\pm$ 0.002 & $\pm$ 0.02 & $\pm$ 0.02 & $\pm$ 100 & $\pm$ 100 & $\pm$ 100 \\
\hline
\textbf{LH} $\bm{\fNLloc}$ & 1000 & 0.834 & 0.3175 & 0.049 & 0.9624 & 0.6711 & [-300, 300] & 0 & 0\\
\textbf{LH} $\bm{\fNLeq}$ & 1000 & [0.6, 1.0] & [0.1, 0.5] & 0.049 & [0.8, 1.2] & [0.5, 0.9] & 0 & [-600, 600] & 0\\
\textbf{LH} \Quijote & 2000 & [0.6, 1.0] & [0.1, 0.5] & [0.03, 0.07] & [0.8, 1.2] & [0.5, 0.9] & 0 & 0 & 0\\
\enddata
\end{deluxetable*}

\vspace{-15px}
\section{Statistics}
\label{sec:statistics}

In this work, we test the performance of a variety of summary statistics calculated on the \QuijotePNG\ halo catalogues in redshift-space.

First, we consider the two and three-point correlation functions of the halo density field in Fourier space $\delta(\bm{k})$, namely the power spectrum and bispectrum. We use the same estimation pipeline as in \citet{Jung:2022gfa, Jung:2023kjh}. The halo power spectrum is estimated up to $\kmax=0.5\,\hMpc$, using bins of size $k_f$ (fundamental mode of the grid) and considering halos of mass above $\Mmin=3.2 \times 10^{13} \, M_\odot/h$. The halo bispectrum is computed up to the same non-linear scales, using a modal estimator \citep{Fergusson:2009nv, Fergusson:2010dm, Fergusson:2010ia, Schmittfull:2012hq}. It simply consists on fitting well-chosen templates to the data, and it has been shown to be extremely efficient to compress the LSS bispectrum \citep{Byun:2020rgl, Jung:2022gfa}, with respect to a standard binned approach. Only a relatively small number of modes, typically less than 100, based on polynomial and tree-level matter bispectrum functions, are necessary to contain the full bispectral information up to $\kmax=0.5\,\hMpc$.

Second, we study the marked (halo) power spectrum and bispectrum. These marked statistics are computed using the same pipeline as the standard power spectrum and bispectrum above, with an extra initial step of weighting the density field. For this weighting operation, we use the mark of \citet{White:2016yhs}, 
\begin{equation}
    \label{eq:mark}
    m(\bm{x};R,p,\delta_s) = \left[ \frac{1+\delta_s}{1 + \delta_s + \delta_R(\bm{x})}\right]^p,
\end{equation}
where $\delta_R(\bm{x}$) is the local density field, computed by smoothing the density field with a top-hat filter at scale $R$. The two other parameters, $\delta_s$ and $p$, can be chosen to modify the impact of $\delta_R(\bm{x})$ on the mark (sensitivity and enhancement of low/high density regions). As shown in \citet{Massara:2020pli, Massara:2022zrf}, measuring the power spectrum of such marked density field gives access to new cosmological information with respect to the power spectrum of the standard field, which comes in fact from higher order statistics. Here, we go one step further and also consider the bispectrum of the marked field. 
Due to the larger computational time required to estimate (marked) bispectra than their power spectrum counterparts, we restrict our analysis to the four different marks defined by the following choice of parameters, $R=[30, 25, 20, 30]\,h^{-1}$Mpc, $p=[1, 1, 1, 1]$ and $\delta_s=[0.10, 0.25, 0.50, 0.50]$,
and which were identified in \citet{Massara:2022zrf} as giving the most stringent constraints on cosmological parameters from galaxy catalogues constructed from the \Quijote\ N-body simulations. To construct the density fields of the \QuijotePNG\ simulations, and to compute the corresponding marks, we use the \textsc{Pylians3}\footnote{\url{https://github.com/franciscovillaescusa/Pylians3}} library.

Finally, we also include the halo mass function (HMF) in our analyses, as it was shown to contain significant information about PNG in \citet{Jung:2023kjh}. Here, the HMF is computed using $13$ logarithmic mass bins between approximately $4.0 \times 10^{13}$ and $4.6 \times 10^{15}~M_\odot/h$.

Beyond these different summary statistics that will constitute our baseline analysis in Section~\ref{sec:analysis}, we also use compressed combinations of them calculated with
\begin{equation}
    \label{eq:compression}
    \tilde{s}_i = \left(\frac{\partial \bar{\bs}}{\partial \theta_i}\right)_* \cov^{-1}_* (\bs - \bar{\bs}_*),
\end{equation}
where $\bs$ is a chosen set of summary statistics of expected mean $\bar{\bs}$ and covariance $\cov$, and the subscript $*$ denotes quantities evaluated at a chosen fiducial cosmology. This expression results in one compressed mode $\tilde{s}_i$ per parameter of interest $\theta_i$ and has been shown to be optimal \citep{Heavens:1999am, Alsing:2017var}, i.e.\ preserving the full information about $\btheta$, if the summary statistics follow a Gaussian distribution where only the mean depends on parameters.

In what follows, covariances and derivatives are evaluated at the \Quijote\ fiducial cosmology (see Table~\ref{tab:quijote}). We use the set of $15000$ fiducial simulations, applying the Hartlap correction factor \citep{Hartlap:2006kj} to obtain unbiased estimates of the inverse covariances, and the couples of $500$ displaced simulations to compute derivatives by finite difference.

\section{Methods}
\label{sec:method}

\subsection{Moment network}
\label{sec:NN}

As a way to quickly explore different summary statistics combinations while also covering a wide range of parameters, we train fully connected NNs to perform likelihood-free inference on different summary statistics, using the moment network methodology \citep{Jeffrey:2020itg}.
These NNs will output two numbers, $\hat{\theta}$ and $\hat{\sigma}$ for each target parameter $\theta$; the first being the mean and the second being the standard deviation of the marginalized posterior.
To do so, we use the loss from, e.g., \citet{CAMELS:2021raw}
\begin{equation}\label{eq:logMSE}
\begin{split}
    \mathcal{L}_\text{logMSE}
    =
    \sum_{i \in \text{pars.}} 
&
    \log \left[
        \sum_{j\in \text{batch}}
        \Delta_{i,j}^2
    \right]
\\
    + \sum_{i \in \text{pars.}} 
& 
    \log \left[
         \sum_{j\in \text{batch}}
         \left( \Delta_{i,j}^2 - \hat{\sigma}^2_{i,j} \right)^2
    \right] ,
\end{split}
\end{equation}
where $\Delta_{i,j} \equiv \theta_{i,j} - \hat{\theta}_{i,j}$. The logarithms have been introduced to make both terms on the RHS of the same order of magnitude. Using this loss, it is guaranteed that the output of the network represents the first two moments of the posterior without making assumptions about its shape. Further details about the NNs architecture and training are provided in Appendix \ref{app:NN_architecture}

\subsection{Evaluating performance}
\label{sec:performance_parameters}

In our analysis, we consider several indicators to monitor the quality of the moment predictions.

First of all, for each parameter we calculate the coefficient of determination 
\begin{equation}
    R^2 (\hat{\theta})
    \equiv 1 - 
    \frac{
    \sum_i
    \left(
        \theta_i - \hat{\theta}(x_i)
    \right)^2
    }{
    \sum_i
    \left(
        \theta_i - \bar{\theta}
    \right)^2
    } \, ,
\end{equation}
where $i$ runs over simulations in the test set, $\theta_i$ is the input (true) parameter for the $i$-th simulation, $\bar{\theta}$ is the average of the true parameter over the entire test set and $\hat{\theta}(x_i)$ is the posterior mean estimate, extracted from the $i$-th simulation.
We notice that $R^2=1$ if the true parameters are exactly recovered, whereas $R^2=0$ if the average value is always used as a prediction. Let us also stress that, despite the symbol used, $R^2$ can be negative if the estimator performs worse than just using the average value.

The coefficient of determination informs us of the quality of the posterior mean estimates, while a different metric needs to be used to monitor the second moment.
Separately for each parameter again, we also calculate the coefficient
\begin{equation}
    \chi^2 (\hat{\theta}, \hat{\sigma})
    \equiv
    \frac{1}{N}
    \sum_i
    \frac{
    \left(
        \theta_i - \hat{\theta}(x_i)
    \right)^2
    }{
    \hat{\sigma}^2(x_i)
    } \, ,
\end{equation}
where $\hat{\sigma}(x_i)$ is the standard deviation prediction based on the $i$-th simulation statistics, and $N$ is the number of simulations in the considered set. This estimator is used to characterize the accuracy of the errors: the closer to 1, the more calibrated they are.
For each trained network, we calculate $\chi^2$ using the simulations in the validation set and discard all 
those where
$|\chi^2-1|>0.5$.
Instead, all the instances of $\chi^2$ shown below are calculated using the simulations in the test set.

\section{Results}
\label{sec:analysis}

\subsection{Fisher forecasts}
\label{sec:fisher-results}

We start by evaluating the information content of the different statistics presented in Section~\ref{sec:statistics} by considering the fiducial parameter values summarized in Table~\ref{tab:quijote} and adopting the Fisher matrix formalism. 

We use the combined Fisher estimator of \citet{Coulton:2023sfu} to obtain unbiased results with the limited number of simulations at our disposal. For the details of the implementation, we refer the reader to \citet{Jung:2023kjh}, where we already applied this method to the same halo power spectra, bispectra and HMF.

\begin{figure*}
    \centering
    \includegraphics[trim={0 .1cm 0 0}, width=0.99\linewidth]{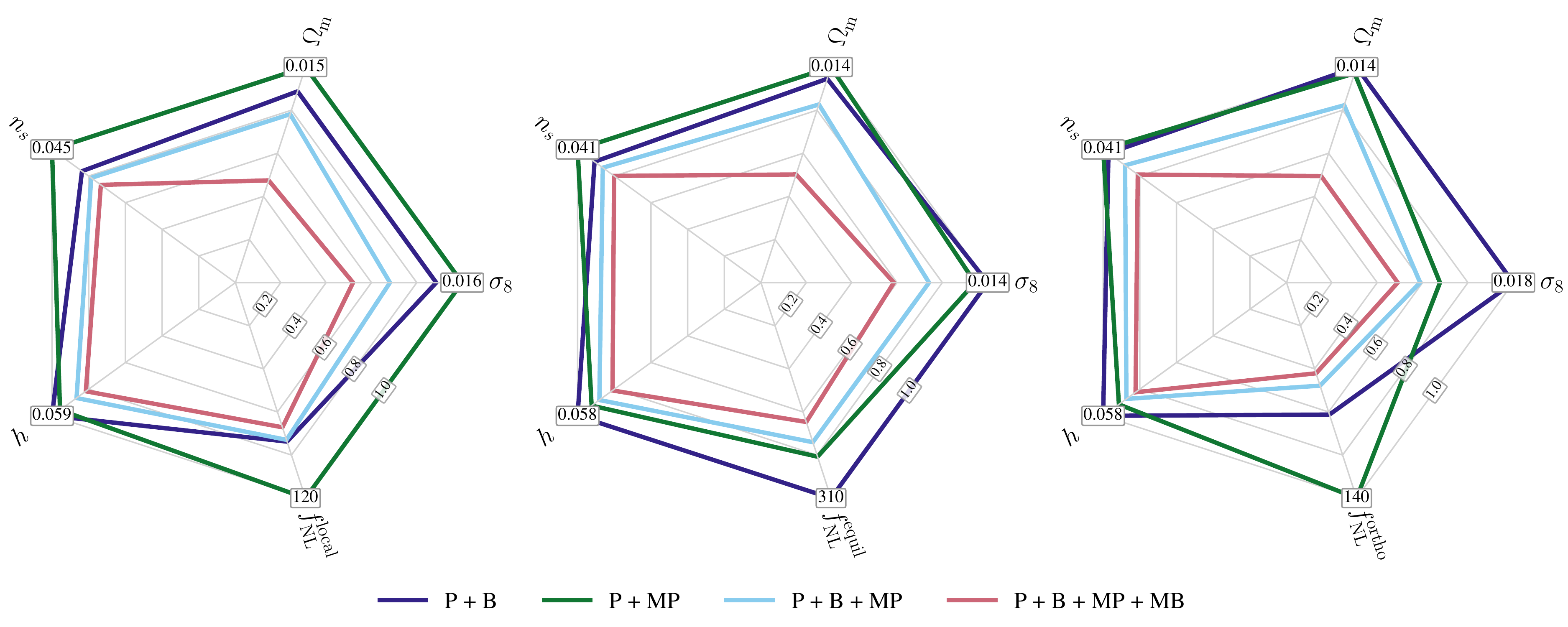}
    \caption{The $1$-$\sigma$ Fisher error bars from different combinations of summary statistics measured in the \Quijote\ halo catalogues at $z=0$, after marginalizing over $\Mmin$. From left to right we consider the local, equilateral or orthogonal PNG shapes jointly with cosmological parameters. The dark blue line correspond to power spectrum + bispectrum (P+B) constraints, the green line to power spectrum + marked power spectrum (P + MP), both are combined to obtain the light blue lines (P + B + MP) and finally the red lines also include the marked bispectrum (P + B + MP + MB). The bold values are the largest $1$-$\sigma$ Fisher error bars obtained for each parameter, used as normalization for the others. Marked statistics contain new information with respect to their standard counterparts on all parameters, which is significant for $\fNLeq$, $\fNLort$, $\sigma_8$ and $\Om$.
}
    \label{fig:fisher}
\end{figure*}

In Figure~\ref{fig:fisher} we compare the $1$-$\sigma$ Fisher error bars on cosmological parameters and PNG amplitudes for different combinations of summary statistics. An important result is that adding the marked power spectrum information to a standard power spectrum and bispectrum analysis improves the constraints on all parameters, except $\fNLloc$. This effect is the strongest for PNG of the equilateral and orthogonal types (more than $20\%$ decrease), as well as $\sigma_8$ (close to $40\%$). A further gain is possible by including the marked bispectrum as well, where we even see a $10\%$ improvement on $\fNLloc$, almost $50\%$ for $\sigma_8$ and $\Om$, and close to $40\%$ on $\fNLeq$ and $\fNLort$. A similar analysis focusing only on cosmological parameters is presented in appendix~\ref{app:fisher-cosmo}.

\begin{figure*}
    \centering
    \includegraphics[trim={0 .1cm 0 0}, width=0.99\linewidth]{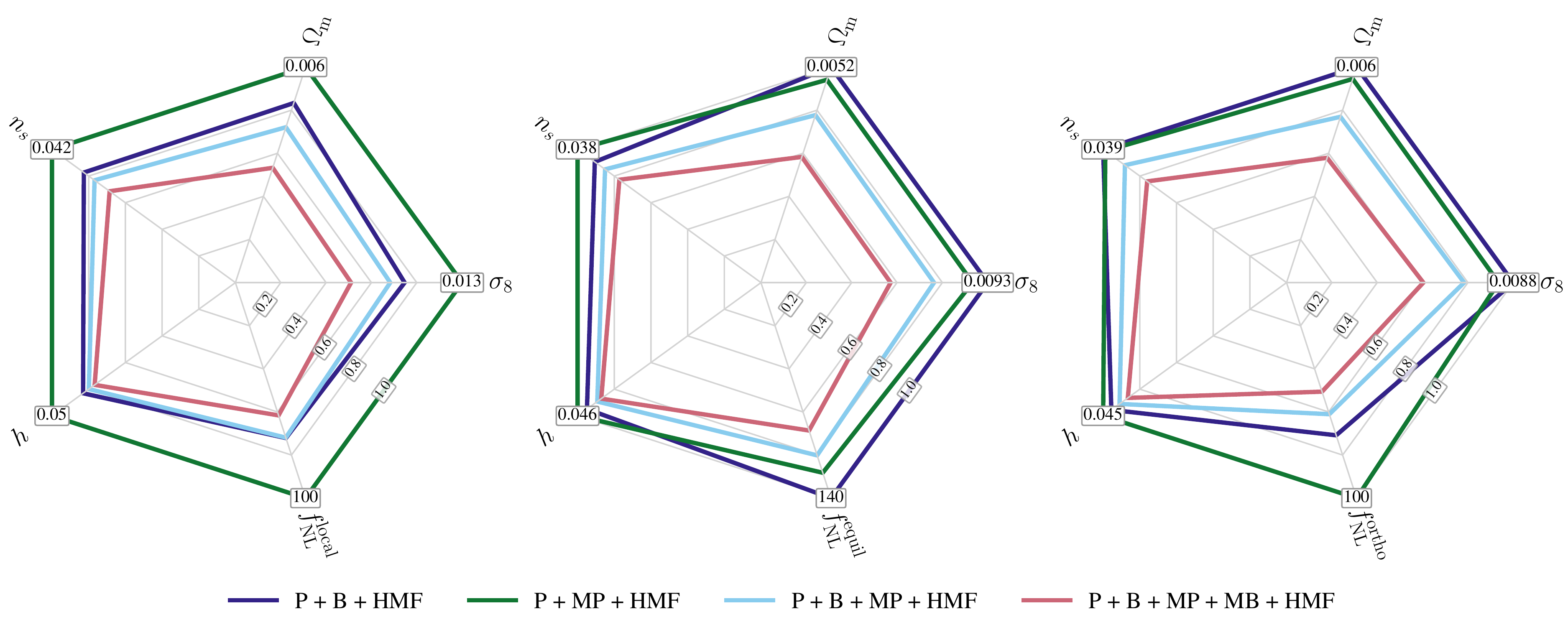}
    \caption{Same as Figure~\ref{fig:fisher}, adding the information of the HMF to every combination of summary statistics, confirming that improvements coming from the HMF or marked statistics have different origins.
    }
    \label{fig:fisher-hmf}
\end{figure*}

In \citet{Jung:2023kjh}, it was shown that the HMF can also bring the same order of improvement on several parameters, especially $\fNLeq$. In Figure~\ref{fig:fisher-hmf}, we verify that these improvements are in fact mostly independent from each other as the HMF and marked statistics bring complementary information. 

Another interesting result shown in Figure~\ref{fig:fisher} is that, in the case of equilateral PNG, a combination of the marked power spectrum and power spectrum performs slightly better than the standard power spectrum and bispectrum analysis. This is not the case for the other two shapes, noticing however that the results are much better than they would be in a power spectrum only analysis \citep[see for example the comparisons in][]{Jung:2022gfa}, except for $\fNLloc$ itself where the constraint comes from the usual scale-dependent bias term of the power spectrum. This is a good indication that the marked power spectrum is a good alternative to the bispectrum in the search of PNG due to its simplicity of estimation with respect to the bispectrum. A comprehensive analysis should encompass both statistics as they contain supplementary information.
 
In these analyses, we use jointly the four sets of marked statistics, defined by  parameters $\{R, p, \delta_s\}$ given in Section~\ref{sec:statistics}. However, a large part of the improvement obtained by including marked statistics is already present when considering only one choice of mark, with a small dependence of the mark defining parameters. For example, the one with the smallest smoothing scale ($R=20$, $p=1$, $\delta_s=0.5$) gives as good results as the combination of the three others, only slightly below (error bars a few $\%$ larger at most) the case where the four marks are used jointly. This is why in many analyses below, where keeping the number of summary statistics as low as possible is important, we will focus on this specific choice of mark. Note also that in principle, it should be possible to improve the constraints even further by optimizing the parameters of the mark, for example by exploring a wide range of them at the power spectrum level, before estimating the more computationally demanding bispectrum. By comparison to other analyses based on marked power spectrum \citep{Massara:2020pli, Massara:2022zrf}, we do not expect any significant difference with the results reported here.

\subsection{Neural network performance}
\label{sec:network_performance}
\begin{figure*}
    \includegraphics[width=0.99\linewidth]{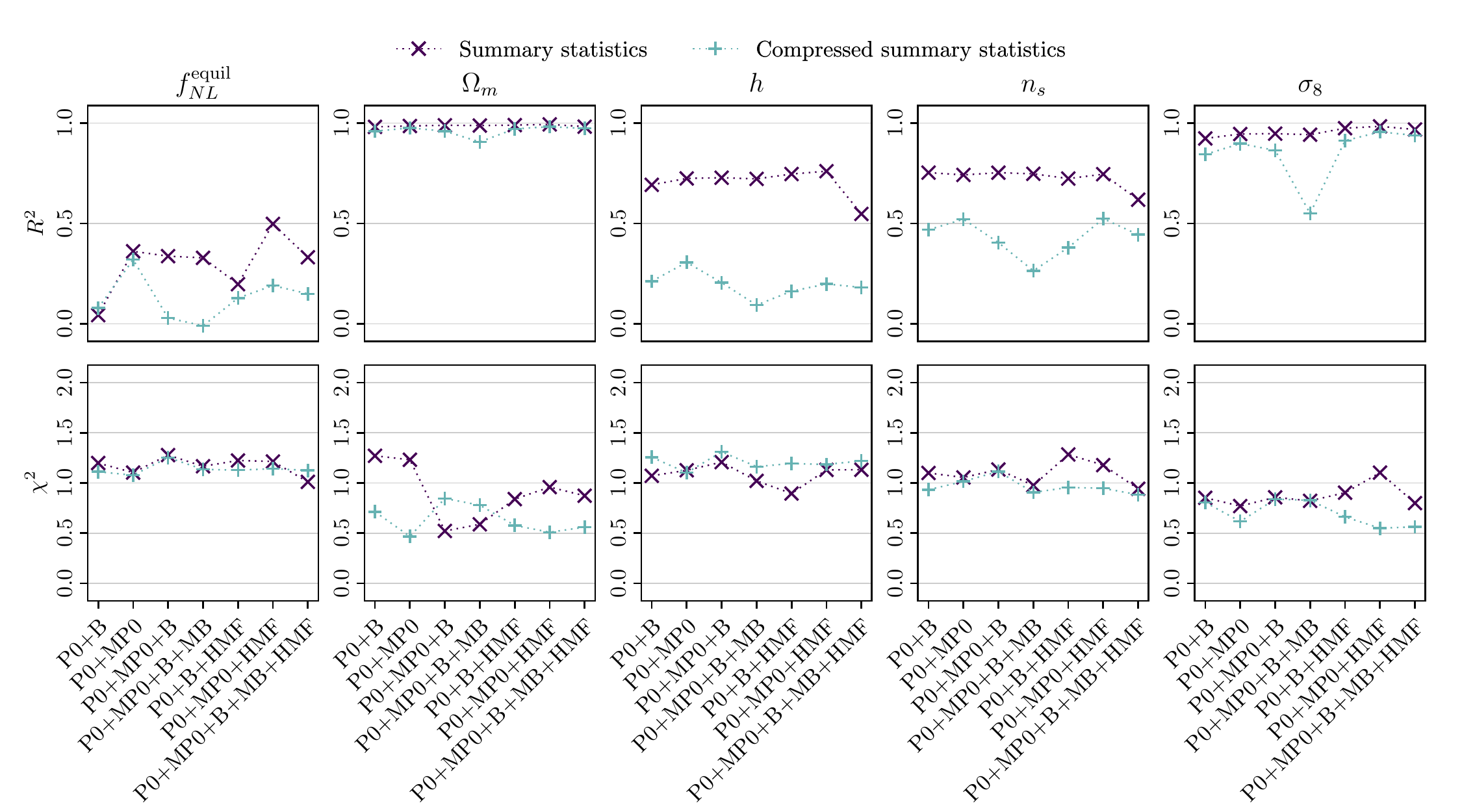}
    \caption{
    Accuracy of the five $\fNLeq$ LH parameter predictions using: power spectrum (P0), marked power spectrum (MP0), bispectrum (B), marked bispectrum (MB) and halo mass function (HMF), each column referring to a specific combination. The input of the NNs are the summary statistics either used as is (purple cross markers) or compressed (light blue plus markers).
    }
    \label{fig:accuracy_comparison_summary}
\end{figure*}
\begin{figure*}
    \includegraphics[width=0.99\linewidth]{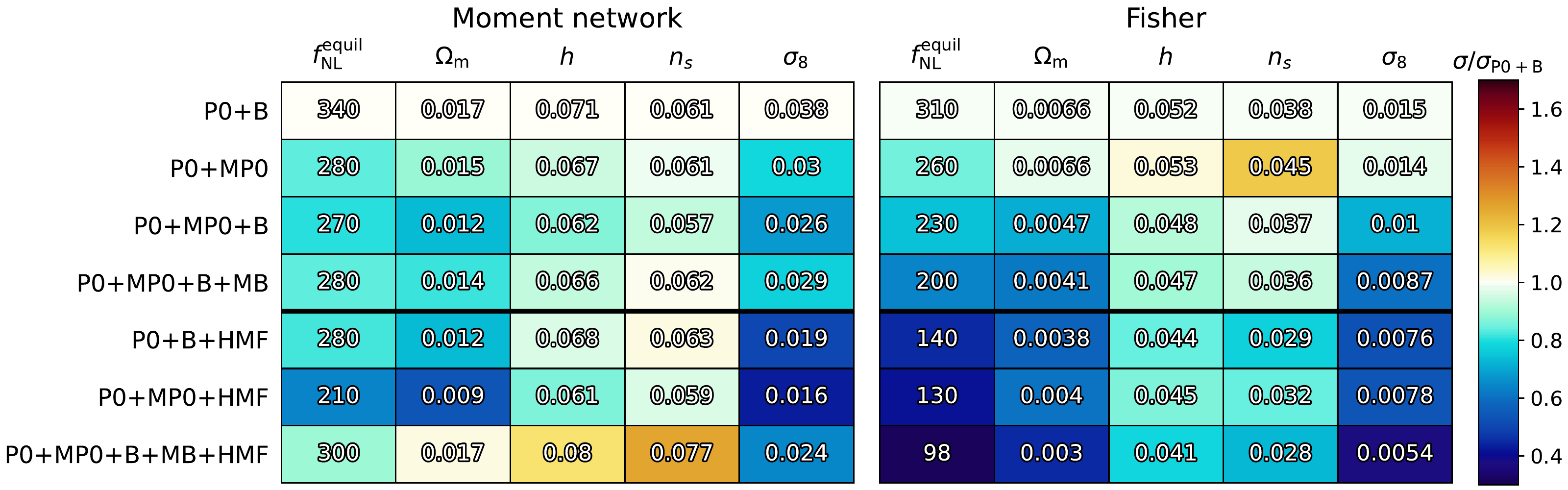}
    \caption{
    Standard deviation predictions for the models in Figure~\ref{fig:accuracy_comparison_summary} applied to the fiducial simulations, and the corresponding $1$-$\sigma$ Fisher constraints. The colour scale gives the ratio of each error with respect to its equivalent computed by using only the power spectrum and bispectrum, highlighting the improvements for almost every parameter with the other combinations of summary statistics. Note that the parameter ranges defining the Latin-hypercube is included as prior to the Fisher calculations.}
    \label{fig:error_comparison_summary}
\end{figure*}

As mentioned earlier, the main goal of this work is comparing and combining the various summary statistics described in Section~\ref{sec:statistics}, over a wide parameter range, by relying on a likelihood-free inference approach based on NNs.
The next few sections are devoted to illustrate our analysis in detail and to discuss several tests aimed at its validation and interpretation.

Our main results are summarized in Figures~\ref{fig:accuracy_comparison_summary} and \ref{fig:error_comparison_summary}.
Since some of the summary statistics are strongly correlated, and the training set we have at hand is limited in size, it is not surprising to sometimes see drops in accuracy in Figure~\ref{fig:accuracy_comparison_summary} as more observables are added to the analysis. A larger number of input features usually requires bigger NN models, which in turn would require a larger training set. If the features one adds are highly informative, the accuracy of the model can still improve. If the new features are highly correlated with the ones already present, the amount of relevant information added may not counterbalance the worse training, and the net result is a drop in accuracy.
The same applies to the standard deviations shown in Figure~\ref{fig:error_comparison_summary}, which in some instances increase when more summary statistics are added.

In the next three sections, we discuss two different ways of pre-processing the data used as input of the NNs identified by different markers in Figure~\ref{fig:accuracy_comparison_summary} (i.e., extracting the summary statistics that are used as is or after the compression step in Equation~\ref{eq:compression}), comment on the comparison between the Fisher bounds and the NN predictions for the fiducial cosmology shown in Figure~\ref{fig:error_comparison_summary}, and benchmark the performance of different combinations of summary statistics.

\subsubsection{Data pre-processing}
We compare two different approaches to data pre-processing. Namely, in one case we directly feed the summary statistics to the networks, whereas in the other we adopt a pre-compression step. In principle, as discussed in \citet{Alsing:2019xrx}, compressing the data decreases the amount of noise while preserving most of the information. Therefore, the NNs trained on the compressed statistic should require a comparatively smaller training set to reach the same accuracy as one naively trained on all the data. However, for the case of $\fNLeq$---one of the main parameters in our analysis---this leads to underwhelming results.

This is shown in Figure~\ref{fig:accuracy_comparison_summary}, where we compare the network trained on the uncompressed summary statistics and the network trained on the compressed statistics. To explain why the compression hinders the training, we have to remember that it is meant to be performed with the maximum likelihood parameters to be optimal, or iteratively until the maximum likelihood is reached.
Instead, we just calculate it once at the fiducial (near the central values of the ranges of parameters covered in the different LHs). In the case of the $\fNLeq$ LH, where five different parameters are varied, none of the simulation has an input set close to the the fiducial, making the compression substantially sub-optimal every time.

We will come back to the subject of data compression in appendix~\ref{sec:Ntrain_stability} where we will apply the same pipeline to different datasets, to glean some information about how many simulations are needed to properly train the network.

\subsubsection{Comparison with Fisher bounds}
Besides looking at $R^2$ and $\chi^2$, which are quantities calculated from the test set of the Latin-hypercubes (see Section~\ref{sec:performance_parameters}), another useful test consists in comparing directly the standard deviations predicted by the NNs applied to the 15000 fiducial simulations with the Fisher bounds. The corresponding results are shown in Figure~\ref{fig:error_comparison_summary}.
An apparent feature is that, even for the best performing NN, the standard deviations are larger than the Cramer-Rao bound.
This means our estimates are somewhat conservative while at the same time the fact we are consistent with this bound provides a validation test of the NN training methodology.
In order to correctly interpret this result, we need to remember that the NN is aimed at building reliable estimates on the {\em whole} Latin-hypercube and not at minimizing the errors for a specific set of parameters. Adding the fact that we have a relatively limited set of simulations at disposal to train the network (see also the discussion in the next section), the observed sub-optimality is to be expected. We discuss this aspect in appendix~\ref{sec:Ntrain_stability}, and a detailed analysis of the convergence with more simulations
will need to be carried out in the future.
For example, in the slightly different setup of \citet{Tucci:2023bag}, numerical convergence is reached with $\sim 10^4$ simulations.

\begin{figure*}
    \includegraphics[width=0.99\linewidth]{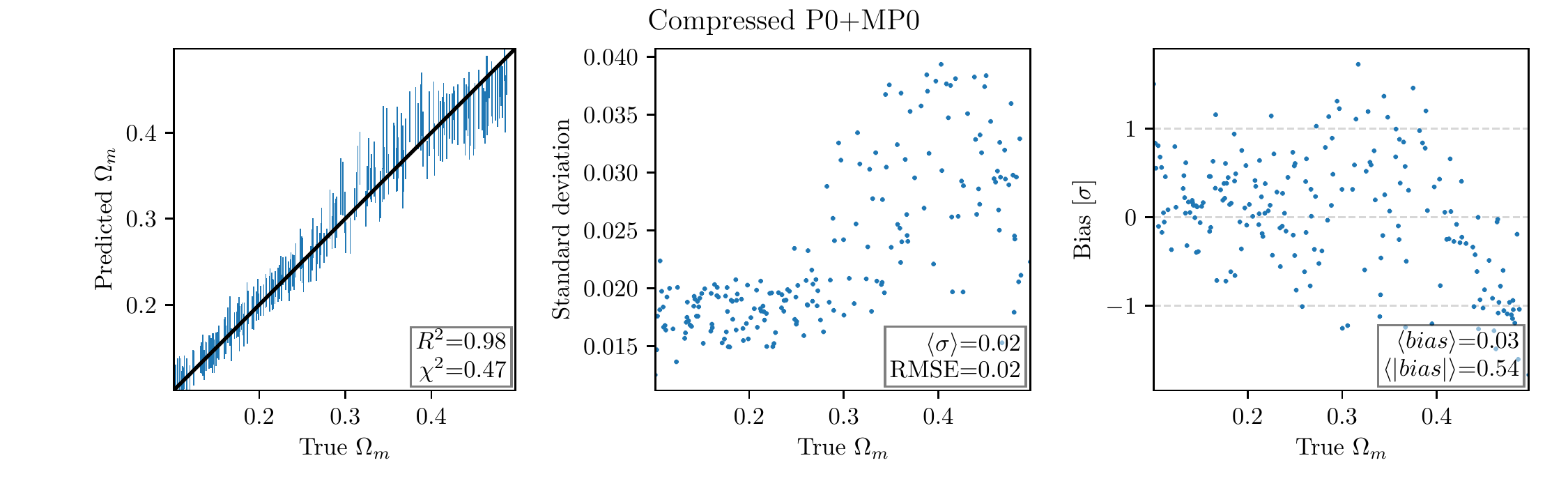}
    \caption{Predictions of $\Om$ of the best performing network trained on the compression of the P0 and MP0 statistics, and applied to the test set of the $\fNLeq$ LH. \emph{Left panel}---Comparison of the true and predicted values of $\Om$. The lines span the predicted 1$\sigma$ about the predicted value.
    \emph{Centre panel}---Predicted standard deviations as a function of the true parameter value. The mean standard deviation is also shown, compared with the root mean squared errors (RMSE) of the mean predictions.
    \emph{Right panel}---Bias of the mean prediction in units of the predicted standard deviations.
    }
    \label{fig:compressed_P0_MP0_Om}
\end{figure*}

The same also holds when NNs are trained on the compressed statistics.
We already discussed how for parameters away from the fiducial, the compression is lossy. As in each simulation multiple parameters, if not all, are displaced from the fiducial, the NN learns to estimate the parameters and their standard deviations from the lossy statistic.
This is combined with the fact that, due to the regularization applied in the training, the NN has to produce a smooth function of the target parameters.
Thus, even when exposed to the compressed statistic calculated in the fiducial, the estimated error bars do not saturate the Cramer-Rao bound. In Figure~\ref{fig:compressed_P0_MP0_Om}, we see how the standard deviations depend on the input parameter, but do not drop in size when close to the fiducial.
We use $\Om$ and the P0 and MP0, as this combination has a high $R^2$ and as such the errors are not driven by the prior.

\subsection{Best combination of observables}
\label{sec:optimal_observables}

In our analysis, we take for granted the use of the power spectrum (P0), which is relevant to constrain the standard cosmological parameters, and assess which other single summary statistic can be added to it to improve constraints on $\fNLeq$. While we show the combination P0+B in Figures~\ref{fig:accuracy_comparison_summary} and \ref{fig:error_comparison_summary} as reference, we find that the marked power spectrum outperforms the bispectrum, when both are complemented with P0. In particular, the mark $R=20$, $p=1$, $\delta_s=0.5$ dominates the others (the figures show the analysis performed with that single marker). These findings are supported by both the Fisher forecasts and the NN analysis.

Adding the bispectrum, alone or with the marked bispectrum, to the analysis of the power spectrum and marked power spectrum shrink the Cramer-Rao bounds, but is quite inconsequential on the NN analysis. This is due to the strong correlation between the two sets of statistics.

Until now we refrained from discussing the inclusion of the HMF, as observationally speaking it may pose additional problems compared to the spectra. However, adding the HMF to any other combination of observables significantly improves both the accuracy and precision of our constraints.
In fact, out of all possible combinations of observables, P0+MP0+HMF is the best performing one; while according to the Fisher forecast an extra $\sim 25\%$ can be gained adding standard and marked bispectra.

\begin{figure*}
    \includegraphics[width=0.99\linewidth]{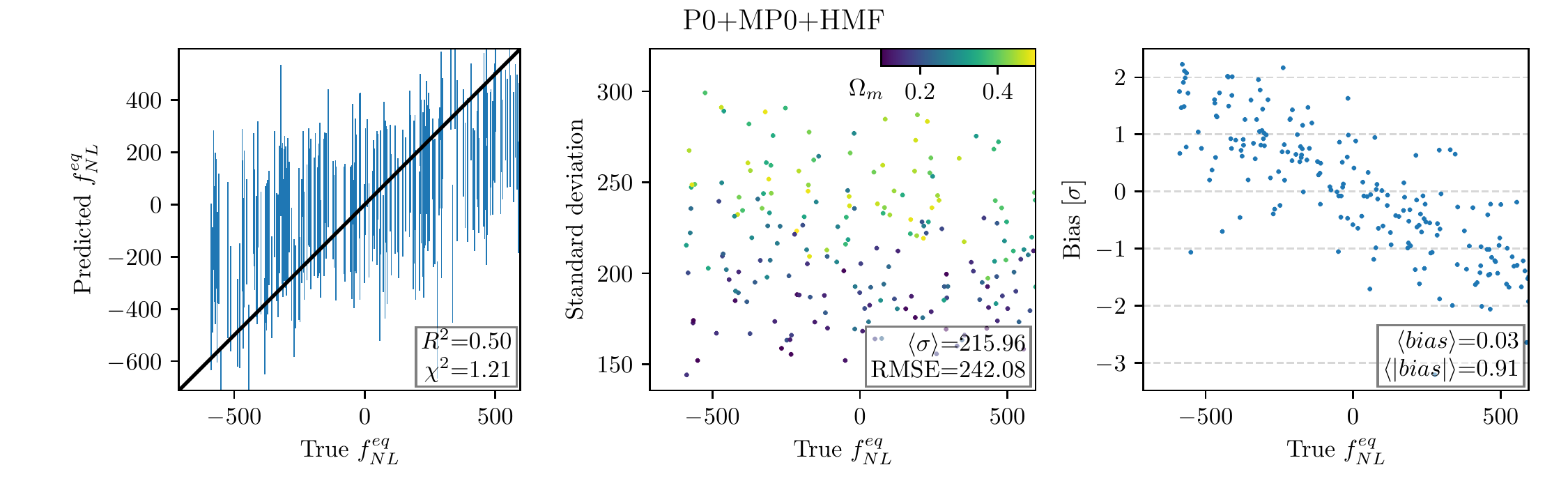}
    \caption{Predictions of $\fNLeq$ of the best performing network trained on P0, MP0 and HMF, and applied to the test set of the $\fNLeq$ LH. The color scale in the middle panel corresponds to the true value of $\Om$, which is the cosmological parameter most correlated to the $\fNLeq$ standard deviation. See Figure~\ref{fig:compressed_P0_MP0_Om} for more details.}
    \label{fig:P0_MP0_HMF_fNLeq}
\end{figure*}

In Figure~\ref{fig:P0_MP0_HMF_fNLeq}, we show the $\fNLeq$ prediction drawn by the NN trained on P0, MP0 and HMF applied to the test set of the $\fNLeq$ LH. Notice that the results are still bounded by the prior. This explains the trend of the bias as a function of the true value of $\fNLeq$. The scatter in the standard deviations is much more sizeable for $\fNLeq$ than for $\Om$ (see Figure~\ref{fig:compressed_P0_MP0_Om}). However, this effect is due to the value of the other cosmological parameters in each simulation, the strongest correlation being with $\Om$ (correlation coefficient of $0.84$).

\section{Conclusion}
\label{sec:conclusion}

In this work we systematically and quantitatively test and rank a variety of summary statistics, power spectrum, bispectrum, their marked counterparts, and the halo mass function, to plan how to optimally analyse LSS data to measure primordial non-Gaussianity.

We achieve this with two complementary approaches. Our Fisher forecast extends the analysis of \citet{Jung:2023kjh}, by adding marked observables for the first time in the PNG context. These forecasts are interesting as they show us the optimal errors we could achieve.
While extra care has been taken to ensure numerical stability, using an additional data compression step, the results
are bound by the choice of fiducial cosmology used to produce the simulations and rely on the assumption of Gaussian likelihood; a limitation that also applies to the quasi-maximum likelihood estimators that can be built with the same pipeline.
These shortcomings are covered by using NNs trained on Latin-hypercubes to perform simulation-based inference. 
They allow us to construct likelihood-free estimators for each summary statistic combination from a relatively low number of simulations (here, an order of magnitude less than the number needed for fully converged Fisher forecast). Moreover, these estimators are reliable on a wide range of parameter values, thus freeing us from the choice of a specific fiducial cosmology.

We choose to use a standard moment network to estimate the marginalized posterior means and standard deviations, and we tested how using the raw summary statistics is better choice than their score-compression, due to computational limitation---the compression would require a new batch of simulations in many points, if not each, of the Latin-hypercube.

From the physical point of view, our main finding is that marked statistics show a great potential for the search of PNG in upcoming LSS data. Both the marked power spectrum and marked bispectrum helps to break degeneracies between PNG amplitudes and cosmological parameters, which are present at the standard power spectrum and bispectrum level, and thus decrease significantly Fisher error bars on all parameters. In addition, the marked power spectrum sets a tight constraint on $\fNLeq$, which outperforms the bispectrum when both are used in conjunction with the power spectrum---we confirm this finding with the analysis of the Latin-hypercube. The same conclusions apply when the halo mass function information is added to the different combinations of summary statistics.

Using the moment network method, we predict $\sigma(\fNLeq) = 280$ from the power spectrum and marked power spectrum measured up to $\kmax=0.5\,\hMpc$ in a volume of $1\,h^{-3}\mathrm{Gpc}^3$, and $\sigma(\fNLeq) = 214$ adding the halo mass function.
If we naively scale the moment network errors with the square root of the volume, we obtain $\sigma(\fNLeq) = 36$ and $28$ on a volume of $60\,h^{-3}\mathrm{Gpc}^3$.

This work allowed us to set the structure of an analysis pipeline. It will be useful to repeat it on simulations which include visible tracers of the dark matter halos to appropriately train the moment network, which will open up the possibility of analysing available data.

\section*{Acknowledgements}
\noindent
GJ acknowledges support from the ANR LOCALIZATION project,
grant ANR-21-CE31-0019 / 490702358 of the French Agence Nationale de la Recherche. 
AR acknowledges support from PRIN-MIUR~2020 METE, under contract no. 2020KB33TP.
ML acknowledges support by the MIUR Progetti di Ricerca di Rilevante Interesse Nazionale (PRIN) Bando 2022 - grant 20228RMX4A.

\appendix

\section{Neural network architecture and training}
\label{app:NN_architecture}
The architecture of our NNs consists in a set of fully connected layers.
The input layer is followed by a normalization layer and a variable number of hidden layers, all with the same number of nodes. The output layer concatenates two sets of variables described in a moment.
We use an ELU activation function \citep{2015arXiv151107289C} in all layers beside the output, and apply a dropout \citep{JMLR:v15:srivastava14a} to each hidden layer of the network.
To further regularize the network, we also use weight decay \citep{2017arXiv171105101L} and stop the training if the validation loss does not decrease for 300 epochs, after which the best weights are restored.
As the target parameters are the mean and standard deviation of each parameter posterior, and the latter is a strictly positive quantity, the output layer combines linear activation functions for the means and ELU+1 for the standard deviations.
Weights are initialized according to  the prescription in \citet{2015arXiv150201852H}.
The training is performed by the Adam optimizer \citep{2014arXiv1412.6980K}, with a cyclical learning rate \citep{2015arXiv150601186S}.
The value of various hyper-parameters is set through Bayesian optimization \citep{omalley2019kerastuner} within some parameter range which we verify a posteriori to be wide enough:
the number of hidden layers (in $[1, 8]$), the number of their nodes (in $[8, 2048]$), the dropout rate (in $[0.3, 0.7]$), the weight decay rate (in $[10^{-5}, 10^{-3}]$, with logarithmic sampling), and the base learning rate (in $[10^{-6}, 10^{-2}]$, with logarithmic sampling).

\section{Fisher forecast without PNG}
\label{app:fisher-cosmo}

\begin{figure}
    \includegraphics[width=0.99\columnwidth]{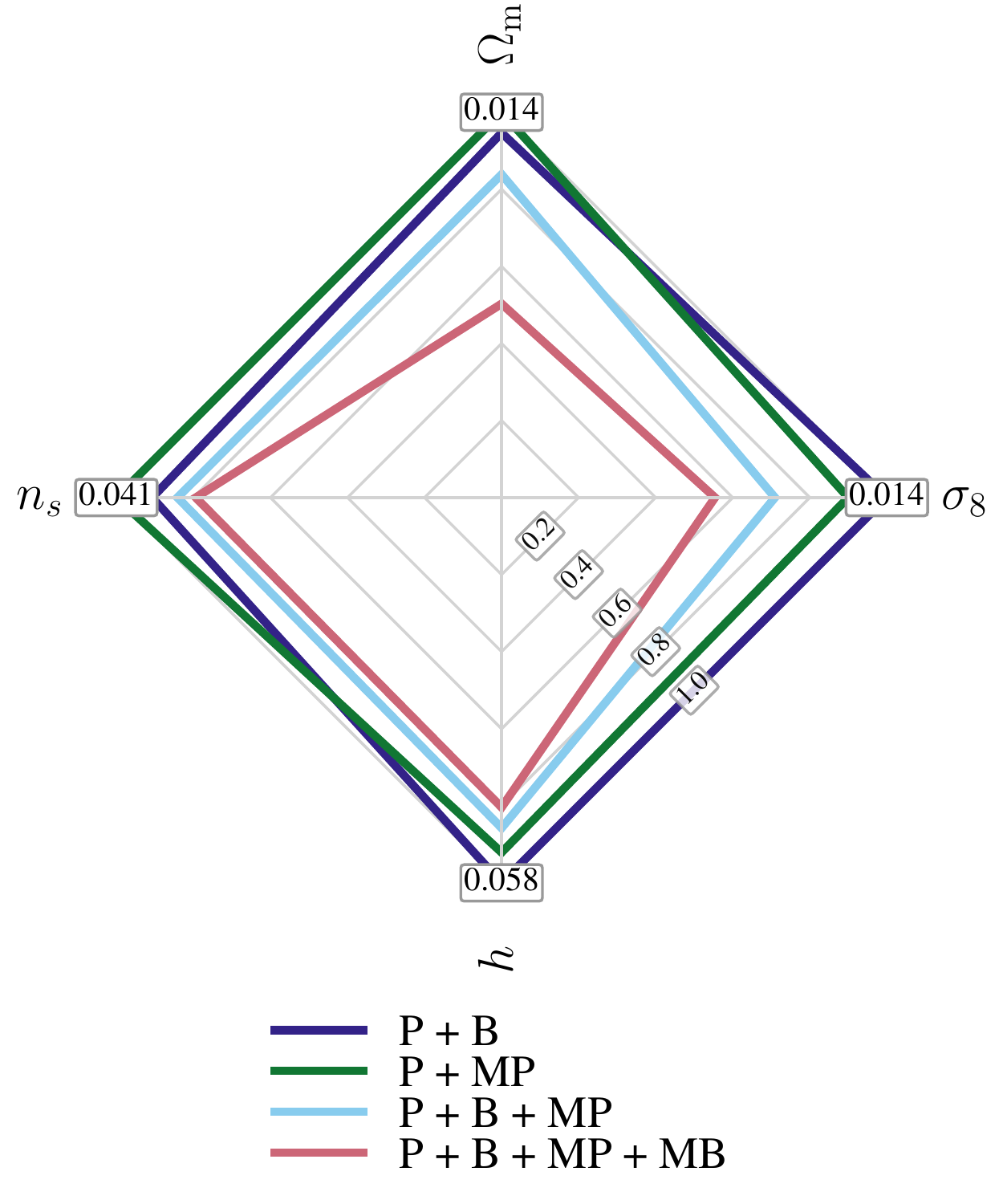}
    \caption{
       The $1$-$\sigma$ Fisher error bars on several cosmological parameters from different combinations of summary statistics measured up to $\kmax=0.5\,\hMpc$ in the \Quijote\ halo catalogues at $z=0$, after marginalizing over $\Mmin$. More details can be found in Figure~\ref{fig:fisher}.
    }
    \label{fig:fisher-cosmo}
\end{figure}

As shown in Section~\ref{sec:fisher-results}, the marked power spectrum and marked bispectrum are powerful observables to study PNG. Here, we verify that they also help in a standard cosmological parameter analysis, as illustrated in Figure~\ref{fig:fisher-cosmo}. While using the power spectrum jointly with bispectrum or marked bispectrum yields similar results, the combination of these three statistics improves the constraints by around $20\%$ for all parameters. Moreover, adding the marked bispectrum helps to disentangle $\sigma_8$ and $\Om$ even further, making the Cramer-Rao bounds of these two parameters around $50\%$ smaller than in the standard analysis.

This confirms that marked observables in general, and particularly the marked bispectrum, are interesting observables for the extraction of cosmological information on non-linear scales.

\section{Small training sample stability}
\label{sec:Ntrain_stability}

\begin{figure}
    \includegraphics[width=0.99\columnwidth]{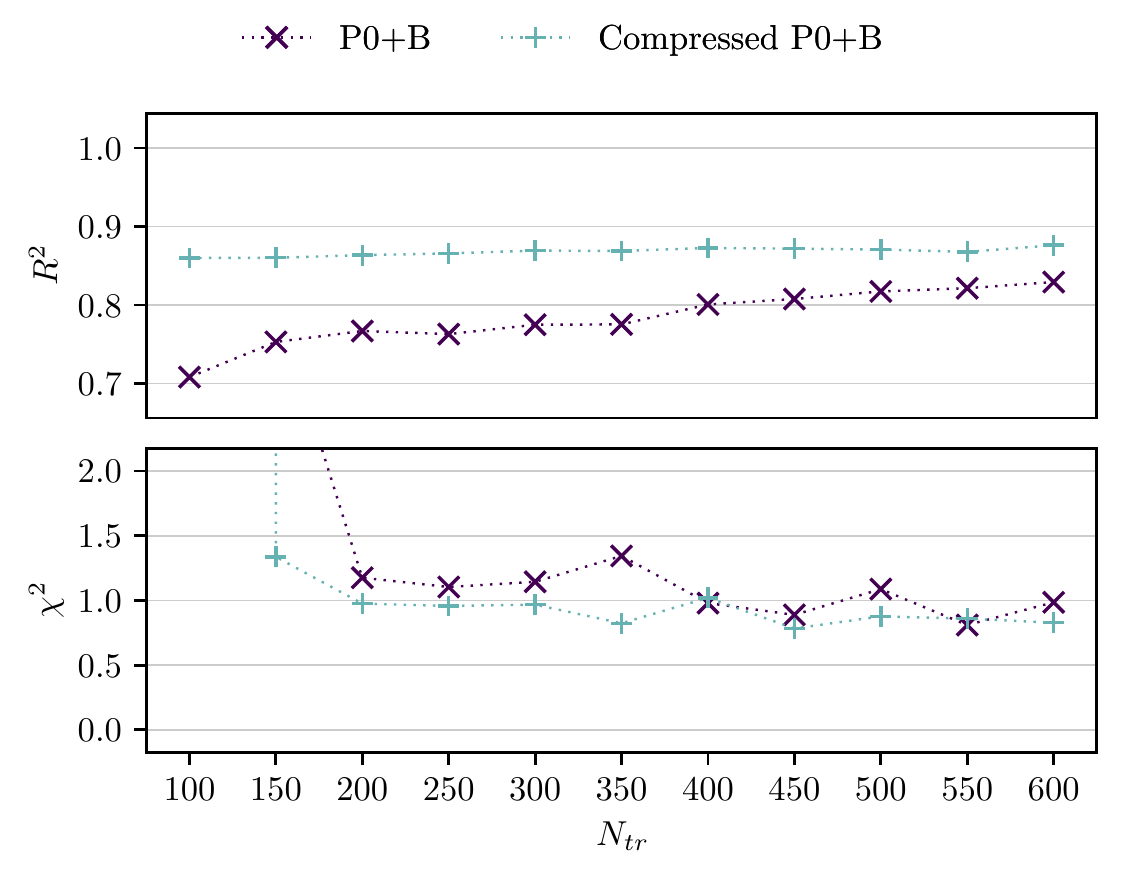}
    \caption{
        Accuracy of the $\fNLloc$ prediction as a function of the number of simulations used to train the model. Compressing the data before training the model simplifies the procedure, which needs comparatively less simulations.
    }
    \label{fig:Accuracy_loc_Ntrain}
\end{figure}

\begin{figure}
    \includegraphics[width=0.99\columnwidth]{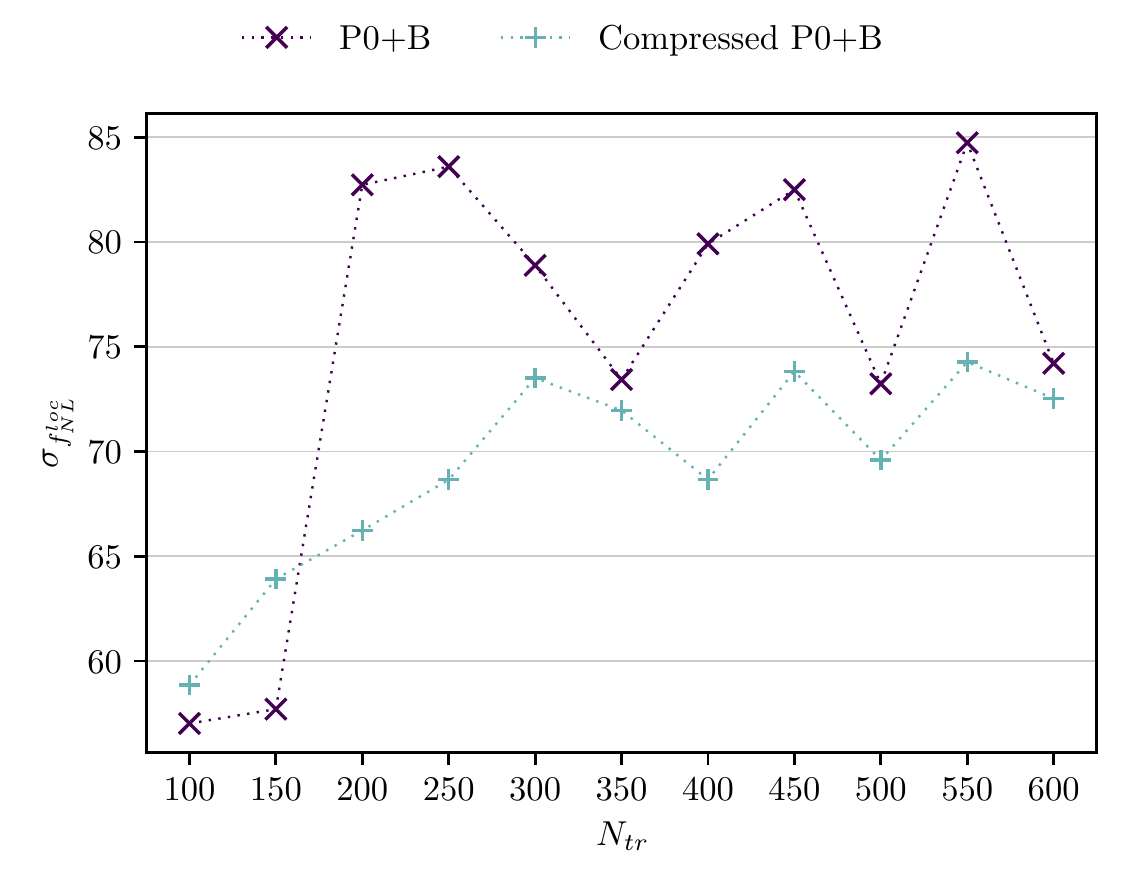}
    \caption{
        Error associated to the $\fNLloc$ prediction as a function of the number of simulations used to train the model. See caption of Figure~\ref{fig:Accuracy_loc_Ntrain}. 
    }
    \label{fig:Error_loc_Ntrain}
\end{figure}

In this appendix, we investigate the convergence of our results with the size of the training sets.
We mentioned in Section~\ref{sec:network_performance} how the compression (if it can be performed correctly) may help to overcome the limits of a small training data set. However, this is not the case for our specific $\fNLeq$ analysis, in which we are limited by the currently available number of simulations, and by the choice of parameters needed to produce the Latin-hypercube in our multi-dimensional parameter space; the same practical issues are present in the vanilla $\Lambda$CDM LH.
Therefore, for the specific purpose of investigating convergence, we choose to study our set of simulations with varying $\fNLloc$ LH. In this case, only one parameter is varied, while leaving all the others fixed at fiducial values. The size of the model (in terms of number of layers and number of nodes per layer) required to fit the data is thus much smaller than in the case of the $\fNLeq$ and of the vanilla $\Lambda$CDM LH sets. Thus, the complexity of the output is lower and the number of training simulations required for the compressed statistics to reach convergence is correspondingly smaller. Even the modest number contained in the PNG-LHs is now sufficient.
In Figure~\ref{fig:Accuracy_loc_Ntrain} we show as an example the $R^2$ and $\chi^2$ values of two sets of NNs trained on the power spectrum and bispectrum, for a variety of number of simulations in the training set $N_{tr}$.
Only in this case, we do not remove the models which have a high $\chi^2$, as it is the matter of the current discussion, but rather take the best model according to the validation loss.
We can see how, without compression, $R^2$ increases throughout the tested range (from $N_{tr}=100$ to $N_{tr}=600$), whereas using the compressed statistic, a couple of hundreds of simulations are enough to give the best achievable estimate with these observables.
For both the compressed and uncompressed statistic, more simulations are required to accurately estimate the error than to estimate the parameter, as shown in the $\chi^2$ panel, and in Figure~\ref{fig:Error_loc_Ntrain}.
However, even in this case, the compressed statistic outperforms the uncompressed statistic when few simulations are at hand.
In either case, the training would benefit from a larger training set.

\begin{figure*}
    \includegraphics[trim={0 2.cm 0 0}, width=0.99\linewidth]{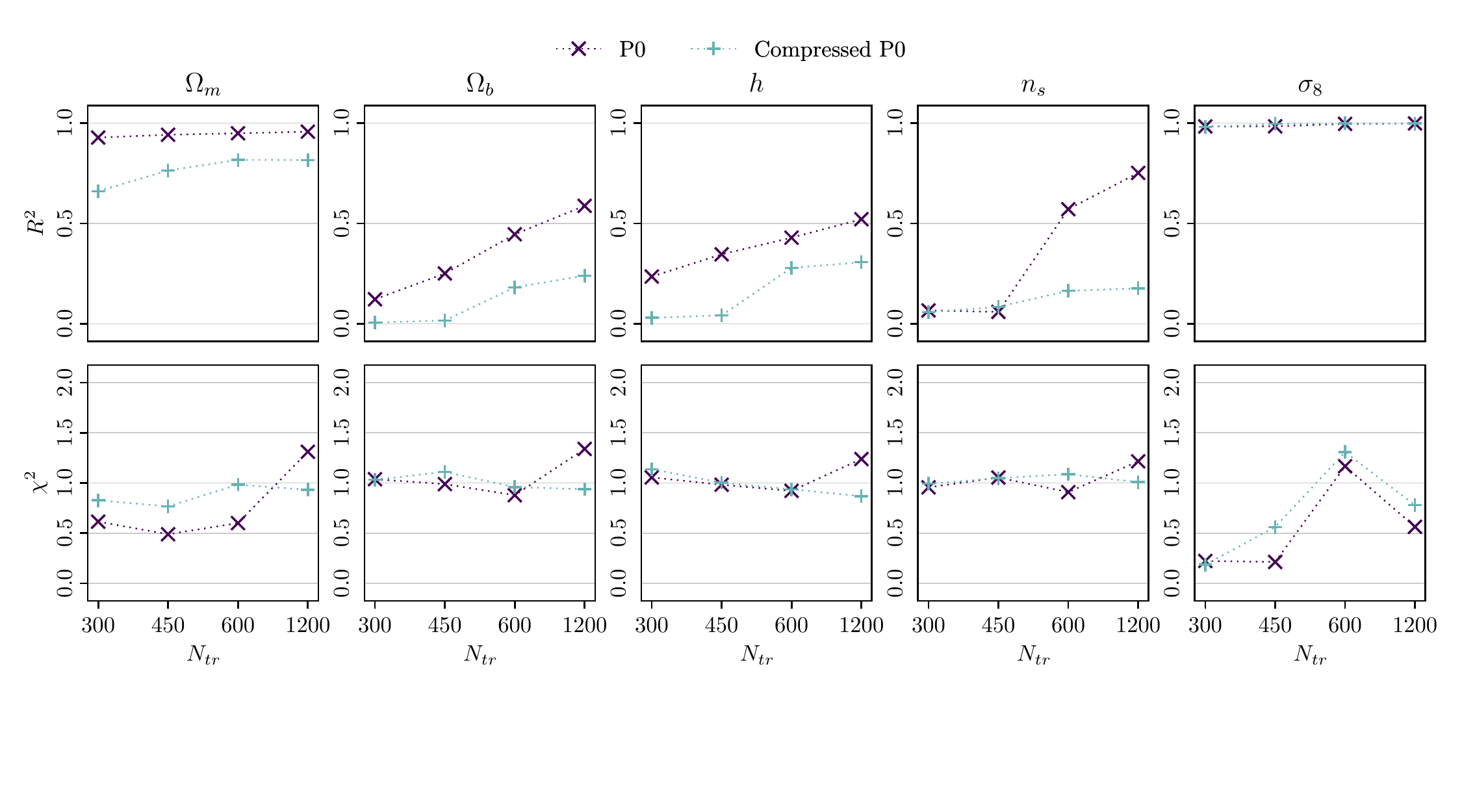}
    \caption{Accuracy of the five vanilla LH parameter predictions analysing either the power spectrum or the compressed power spectrum, as a function of the training sample size.
    Notice that the $x$ axis is not uniformly spaced.}
    \label{fig:Accuracy_van_Ntrain}
\end{figure*}

Despite the fact that the compression leads to unsatisfactory results in the analysis of the vanilla LH, since this case contains twice as many simulations as the $\fNLeq$ LH, it is still interesting to see how the training progresses and the results progressively improve when using a larger and larger training set, as shown in Figure~\ref{fig:Accuracy_van_Ntrain}.

\bibliographystyle{aasjournal}
\bibliography{biblio}

\end{document}